\documentclass[11pt]{article}
\usepackage{latexsym}

\usepackage{amssymb}
\usepackage{amsmath}

 \usepackage{epsfig}
 \usepackage{graphicx}

\usepackage{color}

\numberwithin{equation}{section}  

\newcommand{\be}{\begin{equation}}
\newcommand{\ee}{\end{equation}}
\newcommand{\ba}{\begin{eqnarray}}
\newcommand{\ea}{\end{eqnarray}}
\newcommand{\bi}{\begin{itemize}}
\newcommand{\ei}{\end{itemize}}
\newcommand{\baa}{\begin{array}}
\newcommand{\eaa}{\end{array}}
\newcommand{\edoc}{\end{document}}

\newcommand{\nn}{\nonumber \\ }
\newcommand{\nr}[1]{(\ref{#1})}
\newcommand{\la}[1]{\label{#1}}

\newcommand{\rmi}[1]{{\mbox{\scriptsize #1}}}
\newcommand{\fr}[2]{{\frac{#1}{#2}\,}}
\newcommand{\fra}[2]{{\textstyle{\frac{#1}{#2}\,}}}  

\newcommand{\bfx}{{\bf x}}

\newcommand{\lambdalow}{\Lambda_\rmi{TC}}
\newcommand{\lambdahigh}{\Lambda_\rmi{ETC}}

\newcommand{\Tlow}{T_\rmi{TC}}
\newcommand{\Thigh}{T_\rmi{ETC}}

\def\re{{\rm Re\,}}
\def\im{{\rm Im\,}}

\def\CL{{\cal L}}

\def\CO{{\cal O}}
\def\gsim{\raise0.3ex\hbox{$>$\kern-0.75em\raise-1.1ex\hbox{$\sim$}}}
\def\lsim{\,\,\raise0.3ex\hbox{$<$\kern-0.75em\raise-1.1ex\hbox{$\sim$}}\,\,}

\begin{document}

\begin{titlepage}
\begin{flushright}
HIP-2011-19/TH\\
\today\\ 
\end{flushright}
\begin{centering}
\vfill

{\Large{\bf Mass spectrum and thermodynamics of quasi-conformal
gauge theories from gauge/gravity duality}}

\vspace{0.8cm}

\renewcommand{\thefootnote}{\fnsymbol{footnote}}

J. Alanen$^{\rm a,b}$\footnote{janne.alanen@helsinki.fi},
T. Alho$^{\rm b,c}$\footnote{timo.s.alho@jyu.fi},
K. Kajantie$^{\rm a,b}$\footnote{keijo.kajantie@helsinki.fi},
K. Tuominen$^{\rm b,c}$\footnote{kimmo.i.tuominen@jyu.fi}
\setcounter{footnote}{0}

\vspace{0.8cm}

{\em $^{\rm a}$%
Department of Physics, P.O.Box 64, FI-00014 University of Helsinki,
Finland\\}
{\em $^{\rm b}$%
Helsinki Institute of Physics, P.O.Box 64, FI-00014 University of
Helsinki, Finland\\}
{\em $^{\rm c}$%
Department of
Physics, University of Jyv\"askyl\"a\\}
\vspace*{0.8cm}

\end{centering}

\noindent
We use gauge/gravity duality to study simultaneously the mass spectrum
and the thermodynamics of a generic quasi-conformal gauge theory, specified by its
beta function. The beta function of a quasi-conformal theory almost vanishes, and
the coupling is almost constant between two widely separated energy scales.
Depending on whether the gravity dual has a black hole or not, the mass spectrum
is either a spectrum of quasinormal oscillations or a normal $T=0$ mass spectrum.
The mass spectrum is quantitatively correlated with the thermal properties
of the system. As the theory approaches conformality, the masses
have to vanish. We show that in this limit, the masses calculated via gauge/gravity
duality satisfy expected scaling properties.

\vfill \noindent



\vspace*{1cm}

\noindent

\vfill

\end{titlepage}

\section{Introduction}
In this article, we shall consider gravity dilaton -systems dual to
quasi-conformal gauge theories.
Their characteristic feature is the presence of two widely different energy scales,
$\lambdahigh$ and $\lambdalow$, between
which the coupling constant of the theory evolves very slowly; the theory is
almost conformal, the beta function almost vanishes.
Below $\lambdalow$ and above $\lambdahigh$ the coupling runs similarly
to asymptotically free theories.

Quasi-conformal gauge theories are of
current interest for beyond the Standard Model physics
in the context of walking technicolor \cite{sannino_rev}. Our model lacks
proper flavor dynamics of technicolor, but this connection nevertheless motivates
our notation and the quantitative estimate  
$\lambdahigh\sim 10^3\lambdalow$.

Our gravity dual, Eq.~\nr{ansatz} below,
is based on the improved holographic QCD presented in \cite{kiri2,kiri3,kiri4}
and further developed in \cite{aks,akt,ak}.
The extension of the model from QCD to technicolor is based on building the required
physics in the ansatz for the beta function of the coupling $g^2$ of the theory,
in a certain scheme.
For this we use the ansatz
\be
\beta(\lambda)=-c\lambda^2{(1-\lambda)^2+e\over 1+a\lambda^3},\qquad
\lambda=N_cg^2,
\la{betafn}
\ee
which near $\lambda=0,1$ and $\infty$ can be expanded or approximated as
\ba
\beta(\lambda)
&=&-c(1+e)\,\lambda^2+2c\,\lambda^3-c\,\lambda^4+\CO(\lambda^5),\la{UVexp}\\
&\approx& -{c\over1+a}[(1-\lambda)^2+e],\quad \lambda\,\,{\rm near}\,\,1,\la{betafp}\\
&=&-{c\over a}\,\lambda+{2c\over a}-{c(1+e)\over a}{1\over\lambda}+
\CO({1\over\lambda^2}).
\la{IRexp}
\ea
In these one has built asymptotic freedom in the ultraviolet (UV) domain, i.e., as
$\lambda\to0$, walking near $\lambda=1$ provided $e$ is small and confinement
in the infrared (IR) large $\lambda$ domain if $a=2c/3\,$ \cite{kiri2}.
The values of parameters $c,a,e$ reflect the physics one wants to describe:
$c\gsim10$ fixes the high $T$ phase transition to be of first order \cite{akt}
 and $e$ has to be a small number to describe near conformality.
The case $e=0$
leads to an infrared fixed point (IRFP) \cite{ak}.

Thermodynamics of the model has already been studied in \cite{akt}: the system can exist
in three different phases with phase transitions at $T=\Tlow\approx\lambdalow$ and
$T=\Thigh\approx\lambdahigh$.
The main goal of this article is to study the spectrum of the theory, which is
either a usual $T=0$ mass spectrum, if there is no black hole in the gravity dual or
a quasinormal spectrum \cite{landsteiner} with complex energies, if there is one.
The relation between the numerical
values of $\Tlow,\,\Thigh$ and the mass values is then also obtained.

Of particular interest will be the behavior of the mass spectrum when
conformality is approached, $e\to0$. We shall show that in this limit
the mass spectrum as computed from gauge/gravity duality satisfies
Miransky scaling \cite{conflost}
\be
m(c,e)=A(c,e)\exp\left[-\left(\fr23+{1\over c}\right){\pi\over\sqrt{e}}\right],
\la{mKT}
\ee
where $A(c,e)$ is slowly varying when $e\to0$.

The bottom-up gravity dual presented here lacks detailed flavor
dynamics and is thus not intended to be holographic description of walking
technicolor. Examples of
top down technicolor duals are given in \cite{piai1,piai2}.
In any case its merit is that it presents a concrete gravitational dual with
phase transitions at two widely separated energy scales. A general discussion of these in
terms of multiple big black holes is given in Section G of \cite{kiri3}.

Section 2 discusses the gravity dual and, in particular, the determination of the
dilaton potential which, after solving the Einstein equations, leads to the
beta function \nr{betafn}. Details of numerical evaluation of the background
are also presented. Section 3 reviews the thermodynamics, already discussed in
\cite{akt}. Depending on parameters, the model can describe either first order
or continuous transitions. General principles of determining the masses and
quasinormal modes from Green's functions are summarised in Section 4. One practical
limitation of the present dual is that it can only be computed numerically. However,
one can present a simplified version of the model containing qualitatively its
essential properties \cite{kkvv}, this is applied to the determination of the
quasinormal modes in Section 5. The full computation of the spectrum is carried out
in Section 6.

\section{Gravity dual}
\subsection{Equations}
The gravity equations of the model are as follows \cite{kiri3,akt}.
The gravity action, including the scalar field $\phi(z)$, in standard notation
and in the Einstein frame, is
\be
S={1\over16\pi G_5}\int d^5x\,\sqrt{-g}\left[R-\fra43(\partial_\mu\phi)^2+V(\phi)\right].
\la{Eframeaction}
\ee
With the metric ansatz
\be
ds^2=b^2(z)\left[-f(z)dt^2+d\bfx^2+{dz^2\over f(z)}\right],
\la{ansatz}
\ee
the three functions
$b(z), f(z)$ in the metric and the scalar field $\phi(z)$
are determined from the three equations ($\dot b\equiv b'(z)$, etc.)
\ba
&&6{\dot b^2\over b^2}+3{\ddot b\over b}+3{\dot b\over b}{\dot f\over f}
={b^2\over f}V(\phi),
\label{eq1}\\
&& 6{\dot b^2\over b^2}-3{\ddot b\over b}={\fra43} \dot\phi^2,\label{eq2}\\
&&{\ddot f\over \dot f}+3{\dot b\over b}=0.\label{eq3}
\ea
Further, from the functions so evaluated, the beta function follows as
\be
\beta(\lambda)=b{d\lambda\over db}=b{\dot\lambda\over\dot b},
\quad \lambda(z)  = e^{\phi(z)}\sim  g^2N_c.
\label{crucial}
\ee
Thus $\lambda=e^{\phi(z)}$ is the coupling at the energy scale $b(z)$.
Often it is practical to use the rescaled form
\be
X\equiv {\beta(\lambda)\over3\lambda}
\la{defX}
\ee
and to define
\be
W= -{\dot b\over b^2},\quad b\dot W=2{\dot b^2\over b^2}-{\ddot b\over b},
\la{defW}
\ee
which changes \nr{eq2} to first order form. Inserting further
$\dot W=W'(\lambda)\dot\lambda$ and then $\dot\lambda=\dot b\beta/b=-bW\beta$
permits one to solve $W(\lambda)$ in terms of $\beta$ (Eq.~\nr{Wspecial} below).

Note the two roles, mathematical and physical, played by $\beta(\lambda)$.
Irrespective of the physical
context this combination of $\lambda,\,b$,
arises very naturally in the numerical
solution of \nr{eq1}-\nr{eq3} (see, e.g., \nr{V0} or \nr{W} below).
Physically it models the running of the coupling $\lambda(z)$ as a function
of the energy scale $\sim b(z)$.

Note further that $\beta(\lambda)$ can be solved from \nr{crucial} only after
all the fields have been solved. This requires a knowledge of $V(\lambda)$.
However, given $\beta(\lambda)$ one can
from \nr{eq1} with $f=1$ solve
\be
V_0(\lambda)=12W^2(\lambda)\left[1-\left({\beta\over3\lambda}\right)^2\right],
\quad f(z)=1.
\la{V0}
\ee
Here $W(\lambda)$ can be solved in terms of the beta function \nr{crucial}
and, introducing the explicit ansatz \nr{betafn}, one gets
\ba
{W(\lambda)\over W(0)}&=&\exp\left[-\fra49\int_0^\lambda
d\bar\lambda{\beta(\bar\lambda)
\over\bar\lambda^2}\right]
\la{W}\\
&=&\exp\left\{\fra{2c}{27a}
\left[2 \sqrt{3} a^{1/3} (-2 +
        a^{1/3} (1 + e)) (\arctan[{-1 + 2 a^{1/3} \lambda\over \sqrt{3}}] -
        \arctan{-1\over\sqrt{3}}) + \right.\right.\nn&& \left.\left.\hspace{-2cm}
     a^{1/3} (2 + a^{1/3} (1 + e)) \log[(1 + a^{1/3} \lambda)^2/
     (1 - a^{1/3} \lambda + a^{2/3} \lambda^2)] +
     2 \log[1 + a \lambda^3])\right]\right\},\la{Wspecial}
\ea
where $W(0)$ is the inverse of the AdS radius:
\be
W(0)={1\over\CL}.
\ee

At large $\lambda$, Eq.~\nr{Wspecial} implies that $W(\lambda)\sim\lambda^{4c/(9a)}$
so that $V_0(\lambda)\sim\lambda^{8c/(9a)}$, more detailed
asymptotic expressions are given in \cite{kkvv} and below in Eqs.~\nr{bIR} - \nr{largez}.
As shown in \cite{kiri2}, confinement requires that the large $\lambda$ limit be
$V\sim\lambda^{4/3}(\log\lambda)^{(\alpha-1)/\alpha}$ with $\alpha>1$. Thus we fix
\be
a = \fr23 c.
\ee
A mass spectrum of type $m^2\sim n$ \cite{son1} further implies $\alpha=2$.
This is also built in Eq.~\nr{VIR} below.
To avoid spoiling the UV small-$z$, small-$\lambda$ behavior
we shall use the potential
\be
V(\lambda)=V_0(\lambda)\sqrt{{\log(F+\lambda^4)\over\log F}}.
\la{FF}
\ee
Here $F$ is a parameter which sets the scale at which confinement effects
set in. Since the quasi-conformal effects operate, by construction, at $\lambda\approx1$
we certainly expect $F^{1/4}\gg1$ and choose $F=1000$, to have a large separation
between ETC and TC scales. For comparison,
in \cite{kiri4} the potential for SU($N_c$) Yang-Mills thermodynamics was chosen to be
\be
V_\rmi{IHQCD}(\lambda)={12\over\CL^2}\biggl\{1+0.04128\lambda+(7.35\lambda)^{4/3}
\sqrt{\log[1+(13\lambda)^2]}\biggr\}.
\ee

We re-emphasize that $V(\lambda)$ in \nr{FF} is given in terms of a quantity
$\beta(\lambda)$ which one knows only after one has solved the problem. So one
has to numerically check that the output beta function to sufficient accuracy
reproduces the input one; this seems to work very well \cite{akt}.

\subsection{Numerical solution}
The numerical solution of Eqs. (\ref{eq1}) - ({\ref{eq3}) with $V$ given by
\nr{FF}, means finding a set of
functions $b(z),f(z),\lambda(z)$ parametrised by the
horizon value $\lambda_h=\lambda(z_h)$, $f(z_h)=0$ \cite{kiri4,akt} .
Various physical quantities computed from the set of functions are then
functions of $\lambda_h$. For example, see Fig.~\ref{Tlah1000} for $T=T(\lambda_h)$.

To solve Eqs. (\ref{eq1}) - ({\ref{eq3}) it is practical to use the $W$ in \nr{defW}
to write them in the form
\ba
\dot W&=& 4bW^2-\fra{1}{f}(W\dot f+\fra13 b V),\la{eka}\\
\dot b&=& -b^2W,\la{toka}\\
\dot \lambda&=&\fra32\lambda\sqrt{b\dot W},\la{kolmas}\\
\ddot f&=&3\dot fbW,\la{system}
\ea
Details of numerical integration can be found in Sect. 2.2 of \cite{akt}. We
add here a number of points relevant for the purposes of this article.

\begin{enumerate}
\item The choice of the set of $\lambda_h$ values depends on whether one studies
thermodynamics or the mass spectrum. In the former case one has to
integrate over a large range of temperatures so that one chooses a large
set of $\lambda_h$ values from small values, $\lsim1$ for the UV and
$\gg1$ for the IR. For the mass spectrum one should find a special no-black hole
solution $f(z)=1$. How this is to be done is discussed in Appendix A.
However, one can also possible to start from a $f\not=1$
numerical solution with one very large
value of $\lambda_h$ and correspondingly very large value of $z_h$.
Staying within the values of $0<z<z_h$, for which $f(z)$ is very close
to unity, one has an approximate solution of $b(z),\,\lambda(z)$ for $f=1$.

\item The initial condition $b_h=b(z_h)$ follows from
$b(z_h)=\exp[\int^{\lambda_h}d\lambda/\beta(\lambda)]$:
\ba
c(1+e)\log b_h&=&{1\over\lambda_h}-{2\over 1+e}\log[c(1+e)\lambda_h]
\nn
&&+{1-e+a(1+e)^2\over\sqrt{e}(1+e)}\biggl(\arctan{1-\lambda_h\over\sqrt{e}}-
\arctan{1\over\sqrt{e}}\biggr)\nn
&&-{a(1+e)^2-2\over2(1+e)}\log{(1-\lambda_h)^2+e\over 1+e},
\la{beeinit}
\ea
where the constant of integration is fixed so that the leading UV $\lambda\to0$ behavior is
the 2-loop one, 
i.e., the two last terms in \nr{beeinit} vanish at $\lambda_h=0$.

\item The initial condition $W_h=b_hV(\lambda_h)/(-3\dot f_h)$ follows from \nr{eq1}
by requiring that all fields remain finite at the horizon.
For $\dot f_h$ one can take an arbitrary negative value. The energy unit is
determined so that at $z\to0$
\be
b_0\lambda(z)={1\over\log[1/(\Lambda z)]}-
{b_1\over b_0^2}{\log\log[1/(\Lambda z)]\over\log^2[1/(\Lambda z)]},
\quad b_0=c(1+e),\,\,b_1=-2c,
\la{la2loop}
\ee
where we choose units $\Lambda=1$.

\item Due to large ETC and TC scale separation large values of $z$ enter into the
analysis and it is useful to have an analytic approximation to numerics.
One cannot analytically solve the equations at large $z$ and large $\lambda$, but
one finds that a good approximation for $b(z)$ is
\footnote{
Note that one may use $zb(z)=b_0\exp[-(\Xi z)^2]$
to capture both the UV and IR behavior, as in the model of Section \ref{qnmodel} \cite{kkvv}.
For this discussion of the behavior
deep in the IR, it suffices to take only the Gaussian tail into account.
Formulas for a more general parametrisation $b\sim\exp[-(\Xi z)^\alpha]z^p$
are given in \cite{kkvv}.}
\be
b(z)=b_0 e^{-(\Xi z)^2}.
\ee
$\Xi$ here is an IR scale, analogous to the $\Lambda$ in the UV. Another definition
will be given below in \nr{defXi}.
From the numerical solutions we can determine $\Xi$, and we find that typically
$\Xi/\Lambda\sim0.001$. This is consistent with the large
energy scale splitting ETC/TC$\sim1000$.
Inserting this to the equations of motion above one finds that,
up to corrections $1+\CO(1/z^2)\sim1+\CO(1/\log\lambda)$,
\ba
b(z)&=&b_0 e^{-(\Xi z)^2},\la{bIR}\\
W(z)&=&\fra{2}{\CL}e^{(\Xi z)^2}(\Xi z),
\quad \CL=\fra{b_0}{\Lambda},\la{WzIR}\\
\bar{\lambda}(z)&=&e^{3(\Xi z)^2/2}(\Xi z)^{3/4},\,\,
\Xi z=(\fra23\log\bar{\lambda})^{1/2},\la{laIR}\\
\beta(\lambda)&=&-\fra32\lambda
\left(1+{1\over4(\Xi z)^2}+\CO({1\over z^4})\right),\la{betaIR}\\
V(z)&=&9W^2(z)=\fra{36}{\CL^2}e^{2(\Xi z)^2}(\Xi z)^{2}
=\fra{36}{\CL^2}\bar{\lambda}^{4/3}(\fra23\log\bar{\lambda})^{1/2},\la{VIR}\\
f(z)&=&1-{z_h\over z}e^{3\Xi^2(z^2-z_h^2)},\nn
-\dot f(z_h)&=&4\pi T=6\Xi^2\, z_h.
\la{largez}
\ea
Here $\bar{\lambda}\equiv\lambda/\lambda_0$, where $\lambda_0$ is
the constant of integration. These equations will in practice (Appendix A)
be important as the initial conditions for computing the $f=1$ gravity
solutions.

\item As a concrete example of computed bulk fields, Fig.~\ref{fields} shows
$zb(z),\,\lambda(z),\,W(z)$ for one set of parameters; $\lambda_h=10^6$
with $z_h=1805$ was used here. This also shows
how the required structure with two widely separated energy scales
is reflected in the bulk fields. For example, the coupling $\lambda(z)$ first
approaches an almost stable IR fixed point $\lambda=1$
near $z=1$. However, instead of freezing at this value, $\lambda$ gradually starts
growing and ultimately at large $z$ grows as specified
by equation \nr{laIR} valid at large $z$: $\lambda\sim\exp(\fra32 \Xi^2z^2),
\,W\sim\exp(\Xi^2z^2)$.
The transition between a UV fixed point and a quasi stable IR fixed point is
particularly well seen in the behavior of $b(z)$, which first behaves as $\CL_U/z$,
then as $\CL_I/z$ with a decreasing AdS radius
and ultimately in the far IR shows the gaussian decrease \nr{bIR}.
In fact, the numerically computed $z\,b(z)$ in Fig.~\ref{fields} can be fitted by an
expression of the type
\be
z\,b(z)=\bigl[\CL_U-(\CL_U-\CL_I)\Theta(z-z_\rmi{qir})\bigr]e^{-\Xi^2z^2},
\la{zbfit}
\ee
where in Fig.~\ref{fields} $\CL_U=1.1,\,\CL_I=0.18,\,\Theta(z-z_\rmi{qir})=
\fra12+\fra12\tanh((z-0.6)/0.8), \,\Xi=1/730$.
\end{enumerate}

\begin{figure}[!t]
\begin{center}

\includegraphics[width=0.49\textwidth]{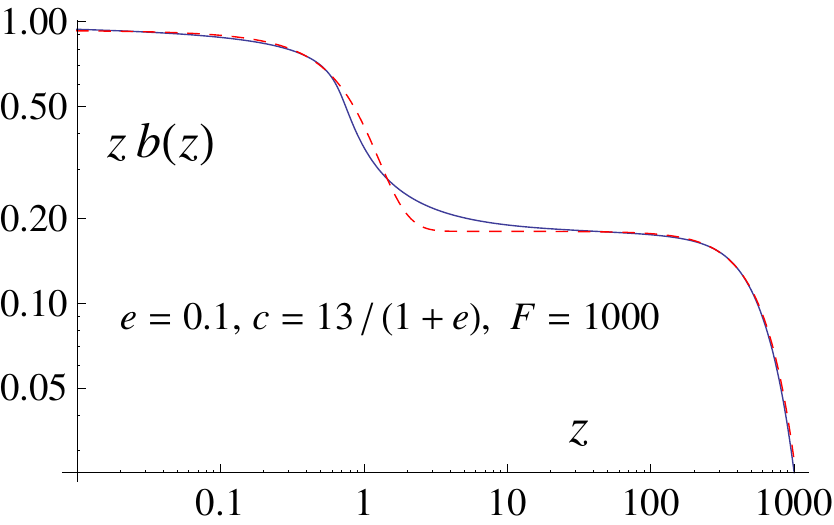}\hfill
\includegraphics[width=0.49\textwidth]{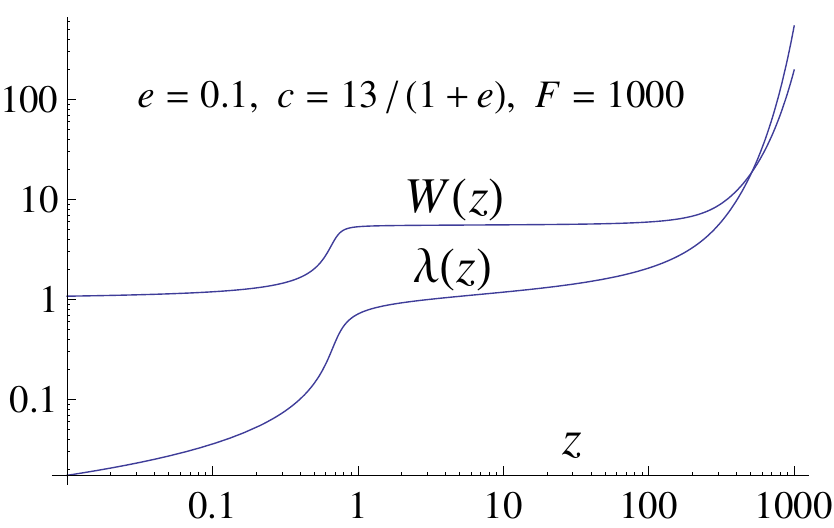}
\end{center}

\caption{\small The functions $zb(z),\,\lambda(z),\,W(z)$ for
$e=0.1$, $c=13/(1+e)$, $F=1000$. The dashed curve is the fit \nr{zbfit} with
parameters given in the text. The horizon is at $z_h=1805$ and at this point
$\lambda(z)$ has grown to the value $10^6$.
}
\la{fields}
\end{figure}

\begin{figure}[!tb]
\begin{center}

\includegraphics[width=0.6\textwidth]{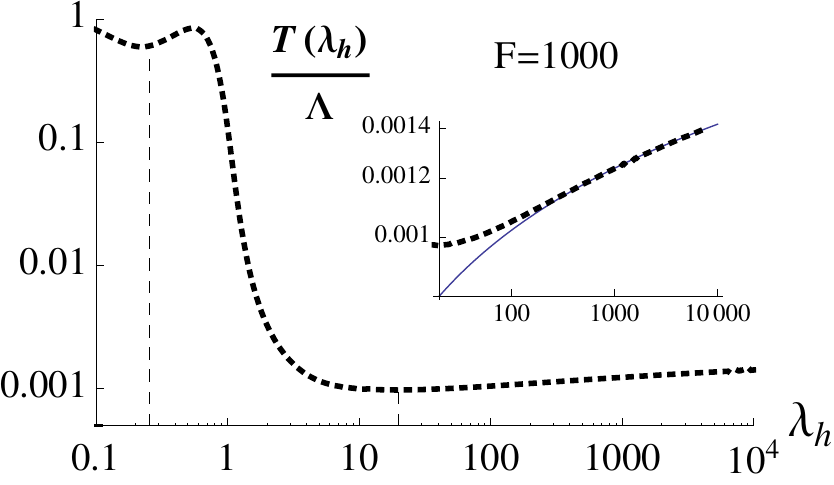}
\end{center}

\caption{\small The function $T(\lambda_h)$ obtained for $e=0.1$,
$c=13/(1+e)$, $F=1000$.
The first minimum is at $\lambda_h=0.218,\,\,T/\Lambda=0.8$ the second at
$\lambda_h=20,\,\,T/\Lambda=0.001$. The corresponding transitions are at
$T/\Lambda=0.653$
(Fig.~\ref{thermoetc}) and at $T/\Lambda=0.00098$ (Fig.~\ref{thermotc}). The
behavior at large $\lambda_h$ (inset) can be fitted by $0.00050 \log\lambda_h^{0.47}$.
}
\la{Tlah1000}
\end{figure}

\begin{figure}[!t]
\begin{center}

\includegraphics[width=0.8\textwidth]{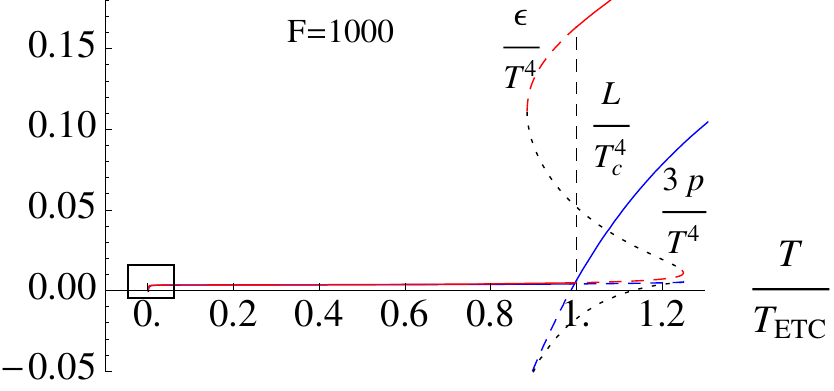}
\end{center}

\caption{\small The equation of state obtained for $e=0.1$, $c=13/(1+e)$,
$F=1000$ for the region
around the
first order ETC phase transition at $\Thigh=0.602\Lambda$. Dotted lines are
unstable, dashed lines
metastable (supercooled or -heated).
The boxed confinement transition to the low $T$ phase with $p=0$ at
$\Tlow=0.000977\Lambda$ is blown
up in Fig.~\ref{thermotc}.
}
\la{thermoetc}
\end{figure}

\begin{figure}[!tb]
\begin{center}

\includegraphics[width=0.8\textwidth]{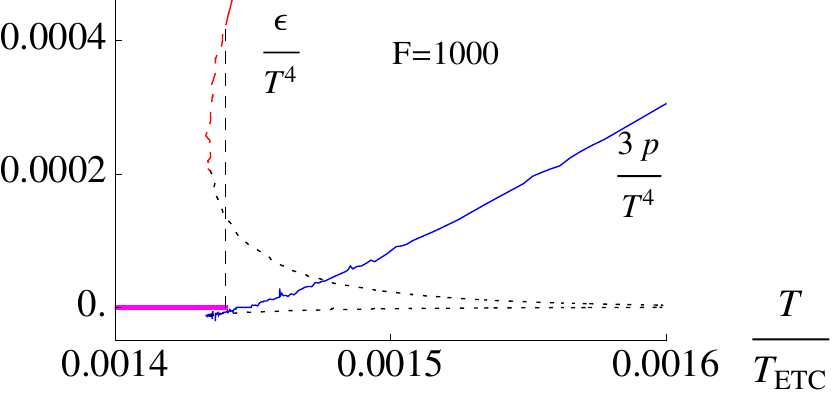}
\end{center}

\caption{\small The equation of state obtained for $e=0.1$, $c=13/(1+e)$,
$F=1000$ in the region
around the confinement transition to a low $T$ phase with $p=0$ (thick red line)
at $\Tlow=0.000977\Lambda$. Dotted lines are unstable, dashed lines
metastable (supercooled or -heated).
}

\la{thermotc}
\end{figure}

\section{Thermodynamics}
The thermodynamics of the present model has been discussed in
\cite{akt}, but with a different functional dependence on the parameter $F$ in
the dilaton potential.
More precisely, in \cite{akt} the $F$-dependent term in \nr{FF} was chosen to
approach one as $e\to0$, while
now the confinement factor in \nr{FF} is independent of the walking
parameter $e$. This is more realistic since confinement enters at large $\lambda$,
far from the IR fixed point $\lambda=1$.

To summarize the results,
we shall choose as reference values
\be
e=0.1,\,c(1+e)=13,\,F=1000.
\la{refval}
\ee
The combination
$c(1+e)$ is the leading 1loop coefficient of the beta function \nr{betafn}.
One then generates numerically a family of solutions $b(z),\,\lambda(z),\,\,f(z)$,
$f(z_h)=0$, parametrised
by the value $\lambda_h=\lambda(z_h)$, and normalises the solutions as in \nr{la2loop}.
From the solutions one then computes thermodynamics by evaluating
\ba
4\pi T&=&-\dot f(z_h),\quad s={b^3(z_h)\over 4G_5}, \\
p&=&{1\over4G_5}\int_{\bar{\lambda}_h}^\infty d\bar{\lambda}_h
\left(-{dT\over d\bar{\lambda}}\right)b^3(\bar{\lambda}_h),
\la{p}\\
\epsilon&=&Ts-p,\\
c_s^2&=&{s(T)\over Ts'(T)}={\beta(\lambda)\over3T}\,{dT\over d\lambda_h}
=\beta(\lambda){\frac{d}{d\lambda_h}}\ln{T^{1/3}}.
\la{cssq}
\ea
Each quantity can be regarded as a function of $z_h$ or $\lambda_h$ or $T$.
If the extreme
UV pressure is $g_\rmi{eff}\pi^2T^4/90$, then
\be
{\CL^3\over4G_5}={2\over45\pi}g_\rmi{eff}.
\ee

Thermal physics is basically contained in the function $T(\lambda_h)$,
shown in Fig. \ref{Tlah1000}. The model is constructed so as to have two
transitions and, accordingly, there are two minima.
The mechanically stable regions with $c_s^2>0$ correspond
to $dT/d\lambda_h<0$ (cf. Eq.~\nr{cssq} and remember that $\beta<0$). With the
structure of Fig.~\ref{Tlah1000} one can have $T$ equal in two different decreasing
branches of $T(\lambda_h)$, and if they also have the same pressure, one has two
different phases in thermal and mechanical equilibrium, i.e. a first order
phase transition, see Fig.~\ref{thermoetc}. The transition temperature $\Thigh$
is somewhat larger than $T$ at the first minimum of Fig.~\ref{Tlah1000}.
In view of the resemblance to
walking technicolor, we call this the extended technicolor transition.

If one continues towards lower temperatures,
at some $T$ the pressure must go to zero, see Fig.~\ref{thermotc}.
It is zero at $\lambda_h=\infty$, as
defined by Eq.~\nr{p}, and since $dT/d\lambda_h>0$, it is negative over
some range in the IR. There is a cusp in $p/T^4$ at the larger minimum
of $T(\lambda_h)$ in Fig.~\ref{Tlah1000}. Below this temperature, $p$ starts
growing and
passes zero at some $T=\Tlow$. The other stable phase is the ``hadron gas''
phase with $p=0$, in this strongly coupled, large $N_c$ theory,
and the transition is the analogue of the deconfining QCD
transition. The large scale separation, $\Thigh/\Tlow\approx600$ is
due to the choice of a large value for the parameter $F$ in \nr{FF}.

A feeling of the numerical values involved, and on the general shape of
$T(\lambda_h)$, can be obtained by adding the UV behavior
$\dot{f}(z_h)\sim 1/z_h$ and the IR limit in \nr{largez} resulting in
\ba
\pi T&=&{1\over z_h}+\fra32\Xi^2\,z_h\nn
&=&\Lambda\, e^{1/(b_0\lambda_h)}(b_0\lambda_h)^{b_1/b_0^2}+
\fra32\Xi\,(\fra23\log\lambda_h)^{1/2}.
\ea
Thus there is a very rapid decrease from the UV and a very slow increase in the IR.
This is also reflected in the numerical value of the IR parameter
$\Xi/\Lambda\sim0.0014$ in \nr{zbfit}. The fit $0.00050 (\log\lambda_h)^{0.47}$ to the
curve in Fig.~\ref{Tlah1000} is compatible with this value of $\Xi$.
We remind that the power $1/2$ of $\log\lambda_h$ was imposed by choosing the
confinement factor $\sim(\log\lambda)^{1/2}$ in \nr{FF}.

\section{Spectrum of the theory: masses and quasinormal modes}
The mass spectrum is obtained as poles of a Green's function of an operator with
appropriate quantum numbers in the background \nr{ansatz} with $f(z)=1$.
The poles corresponding to stable particles are on the real axis;
if the backround has $f(z)$ with a horizon, $f(z_h)=0$, the poles move to
the lower complex $\omega$ plane.

To determine this Green's function in gauge/gravity duality
we need the equation satisfied by fluctuations in the background \nr{ansatz}.
The fluctuations $h_{\mu\nu}$ of the metric satisfying
$\partial_\mu h^{\mu\nu}=h_\mu^\mu=0$, i.e. the tensor fluctuations,
are simple since they do not mix
with other fluctuations. For dilaton
fluctuations, i.e. the scalar fluctuations, one must take into account that they mix
with the fluctuations of the metric. 
The appropriate equation has been derived in
\cite{springer1,springer2}. Keeping first $f\not=1$ and taking $k=0$ the equation
for the gauge invariant scalar fluctuation $\phi\equiv\phi(\omega,z)$ is
\be
\ddot\phi+{d\over dz}\log(fb^3)\cdot\dot\phi-\biggl({\ddot X\over X}+
{d\over dz}\log(f^2b^3)\cdot{\dot X\over X}\biggr)\phi+{\omega^2\over f^2}\phi=0,
\la{spr}
\ee
where $X$ was defined in \nr{defX}. For the tensor fluctuation one simply
removes the $X$-terms. For the true $T=0$ mass spectrum one takes $f=1$ and
replaces $\omega^2\to\omega^2-k^2=m^2$ \cite{kiri_gravfluct}.

The methods of computing the Green's functions and their poles from \nr{spr} are
standard \cite{policastrostarinets,ss,kovtunstarinets,kovtunstarinets2}.
To interpret the outcome it is convenient to transform \nr{spr} to
a Schr\"odinger-like form with $\omega^2$ as the eigenvalue. In order to achieve
this, one removes the $1/f^2$ in \nr{spr} by introducing a new variable
\be
u=\int_0^z {d\bar{z}\over f(\bar{z})},
\la{defu}
\ee
which transforms the interval $0<z<z_h$ to $0<u<\infty$. Introducing further
\be
\psi(u)=\sqrt{b^3}\,\phi(u)
\ee
the equation becomes
\be
-\psi''(u)+V^{(S)}_f(u;z_h)\psi(u)=\omega^2\psi(u),
\la{seq}
\ee
where
\be
V_f^{(S)}(u;z_h)=f^2\biggl[{3\ddot b\over 2b}+{3\dot b^2\over 4b^2}+
{3\dot f\dot b\over 2fb}
+{\ddot X\over X}+\biggl({3\dot b\over b}+{2\dot f\over f}\biggr)
{\dot X\over X}\biggr]_{z=z(u)}.
\la{Vf}
\ee
The superscript $S$ indicates that we are considering scalar fluctuations.
For tensor fluctuation the analysis is similar, and the result for $V_f^{(T)}(u;z_h)$
is obtained from \nr{Vf} by simply removing the $X$-terms.
When $u\approx z\to0$ the requirement $b(z)\to\CL/z$ implies that the potentials
behave as
\be
V_f^{(i)}(u;z_h)\to {15\over4u^2}
\la{VSsmallz}
\ee
and when $u\to\infty,\,\,z\to z_h$,
\be
V_f^{(i)}(u;z_h)\to A^{(i)}(z_h)e^{\dot{f}_h u}\to 0,
\la{Vflimit}
\ee
where
\be
A^{(i)}(z_h) = \left(\frac{d}{dz}{V}_f^{(i)}\vert_{z\to z_h}\right)
\lim_{z_0\rightarrow z_h}(z_h-z_0) e^{-\dot{f}_h u(z_0)}.
\la{VflimitAfactor}
\ee

When $f(z)=1$, the Schr\"odinger-type equation becomes
\be
-\psi''(z)+V^{(i)}_\rmi{Sch}(z)\psi(z)=m^2\psi(z),\quad i=S,T
\la{Seq}
\ee
where
\ba
V_\rmi{Sch}^{(S)}&=&{3\ddot b\over 2b}+{3\dot b^2\over4b^2}+{\ddot X\over X}+
3{\dot b\over b}{\dot X\over X}.
\la{tensscalpot}
\ea
For the tensor Schr\"odinger potential, $V_\rmi{Sch}^{(T)}$, the $X$ terms are omitted.
For a simple estimate of the IR behavior, one may take for $b(z)$ the
numerically accurate fit \nr{zbfit}, $zb(z)=\exp(-\Xi^2z^2)$ valid for large $z$.
This leads to the approximate confining tensor potential
\be
V_\rmi{Sch}^{(T)}={15\over4z^2}+6\Xi^2+9\Xi^4z^2.
\la{VSlargez}
\ee
This has the minimum value $(3\sqrt{15}+6)\Xi^2$ at $\Xi^2 z^2=\sqrt{5/12}$
and the mass spectrum $m_n^2=12\Xi^2(n+2)$.
However, the existence
of the dynamical ETC and TC scales produces interesting structures in between.

\begin{figure}[!t]
\begin{center}

\includegraphics[width=0.49\textwidth]{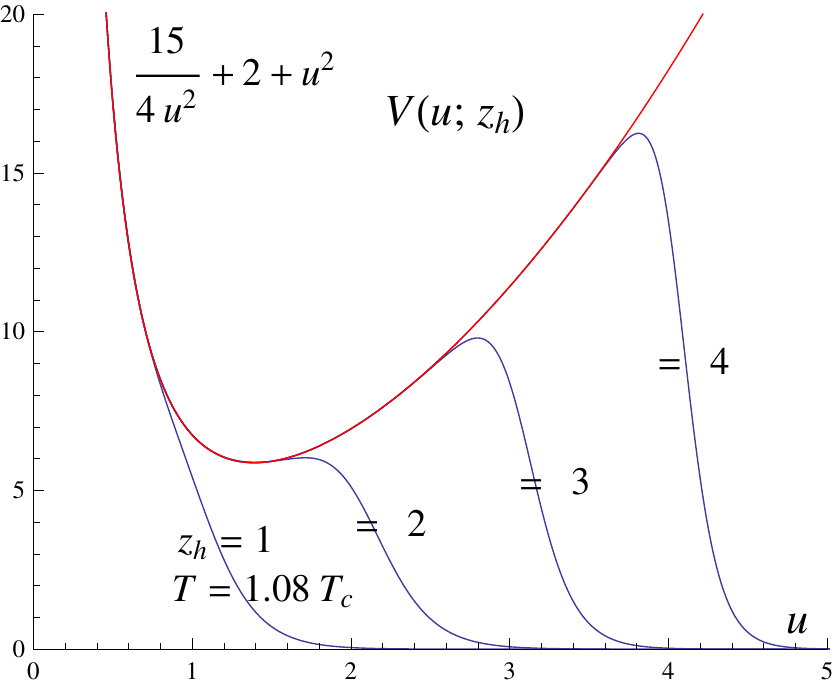}\hfill
\includegraphics[width=0.49\textwidth]{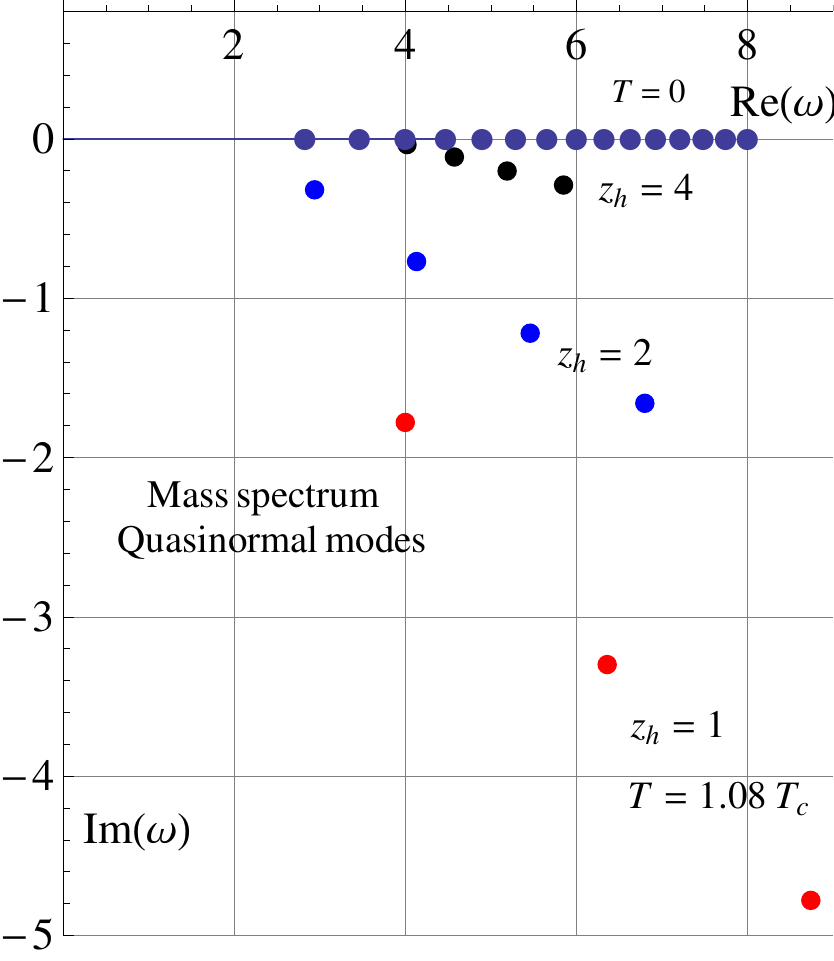}
\end{center}

\caption{\small Left panel: The tensor potential $V_f^{(T)}$ given by \nr{Vf} with
$X$-terms omitted for various values of $z_h$. For
increasing $z_h$ the potential $V_f^{(T)}$ approaches the potential
$V_\rmi{Sch}^{(T)}$ in \nr{tenspot}.
Right panel: The corresponding quasinormal tensor frequences computed from \nr{spr}.
The mass spectrum \nr{gbmass} is plotted on the real axis.
Poles with $\re(\omega)\to-\re\omega$ are not shown.
}
\la{Vffig}
\end{figure}

\section{Spectrum in a model of the deconfinement transition}\la{qnmodel}
To appreciate the relation between $f=1$ and $f\not=1$ cases in the dilaton
background \nr{ansatz}, we shall first consider a simplified version of the
TC or deconfinement transition \cite{kkvv}.
It is based on the ansatz\footnote{For simplicity, we fix the IR scale $\Xi^2=1/3$
in this section.}
\be
b(z)={1\over z}\,e^{-\fra13 z^2},
\la{bz}
\ee
from which one can derive $\lambda(z), \beta(\lambda)$ and, in particular,
\be
f(z)=1-{(z^2-1)e^{z^2}+1\over (z_h^2-1)e^{z_h^2}+1},\quad
T(z_h)={1\over2\pi}\,{z_h^3\over z_h^2-1+e^{-z_h^2}}.
\la{fz}
\ee
The tensor potential in this case is
\be
V_\rmi{Sch}^{(T)}(z)={3\ddot b\over 2b}+{3\dot b^2\over4b^2}={15\over4z^2}+2+z^2,
\la{tenspot}
\ee
and from this the tensor glueball mass spectrum becomes \cite{son1}
\be
m_n=2\sqrt{2+n},\quad n=0,1,2,...\,.
\la{gbmass}
\ee
The curve $T=T(z_h)$ has a minimum $T_\rmi{min}=0.3862$ at $z_h=1.466$
and the system
is in a high $T$ deconfined phase for $T>T_c=0.4000$, $z_h<z_c=1.30$.
Inserting \nr{bz} and \nr{fz} to \nr{tensscalpot} and performing the change of
variables \nr{defu} numerically, one finds the tensor potential plotted
in Fig.~\ref{Vffig}. The overall pattern is obvious: For small $u\approx z$, $f(z)=1$
and $V_f^{(i)}(u;z_h)\approx V_\rmi{Sch}^{(i)}(z)$, both for scalar $(i=S)$ and tensor $(i=T)$.
The effects of $z_h$ start being felt and $f(z)$ starts to decrease towards
$f(z_h)=0$ when $u\approx z$ approaches $z_h$. Furthermore, as $z$ approaches
$z_h$, the variable $u$ approaches infinity and, due to the $f^2(z)$ factor in \nr{Vf},
the potential $V_f^{(i)}(u\to\infty;z_h)$ approaches zero.

The mass spectrum \nr{gbmass} can be derived by solving the Schr\"odinger equation
\nr{Seq}. To solve the quasinormal spectrum when $f\not=1$ it is actually most
convenient to go back to \nr{spr} and use standard techniques for computing the
Green's function. To this end one
solves first Eq. \nr{spr} numerically by starting the integration at $z=z_h$
expanding for small $z-z_h$ so that the leading term
is $\sim (z-z_h)^{i\omega/\dot f(z_h)}$. Secondly, one constructs analytically the
two linearly independent solutions, $\phi_u\sim1$ and $\phi_n\sim z^4$ near $z=0$ by
expanding the solution at small $z$.  Finally, the quasinormal modes
are those values of
$\omega=\re\omega+i\im\omega$ for which the solution $\sim z^4$ at small $z$ and
the coefficient of $\phi_u$ vanishes.
Examples of the lowest quasinormal states, plotted in Fig.~\ref{Vffig}, are
\ba
z_h=1:&&4.0-1.78i,\quad 6.36-3.3i,\quad 8.74-4.78i\nn
z_h=2:&&2.94-0.32i,\quad 4.13-0.77i,\quad 5.46-1.22i,\quad 6.8-1.66i\nn
z_h=4:&&2.83-0.i,\,\, 3.46-0.004i,\,\, 4.02-0.036i,\,\, 4.57-0.11i,
\,\, 5.19-0.20i,\,\,5.85-0.29i\nonumber
\ea
The pattern is clear: for large $z_h$ the quasinormal frequencies approach the real mass
spectrum \nr{gbmass}. However, then the system is in the unstable small black hole
phase with $T'(z_h)>0$. With decreasing $z_h$ the spectrum moves away from the
real axis, and approaches $\omega=2n(\pm1-i)/z_h=2\pi Tn(\pm1-i)$ as the system
enters the stable high $T$ phase. For example, $z_h=1$ corresponds to
$T=1.08T_c$, but then the states are very broad as they are melting away.

\section{Spectrum for quasi-conformal theory}

\begin{figure}[!t]
\begin{center}

\includegraphics[width=0.6\textwidth]{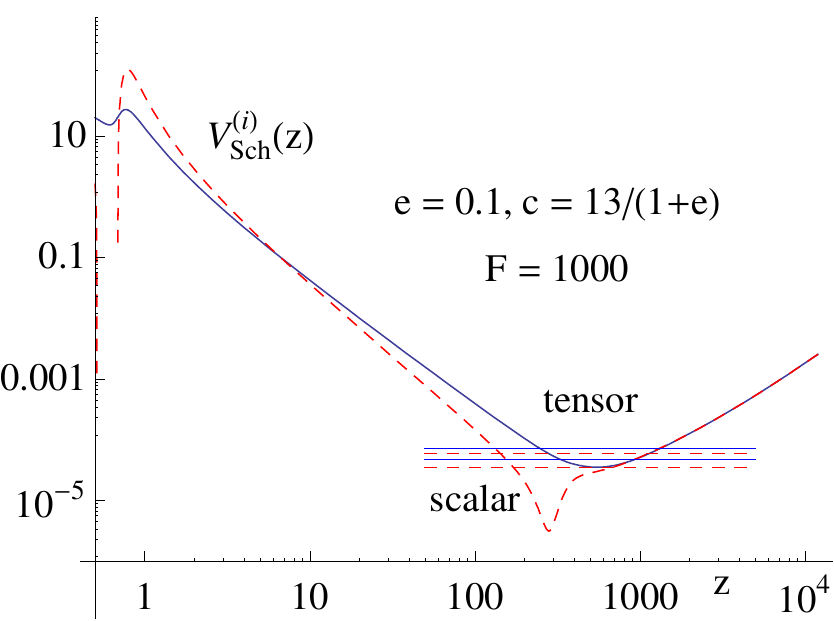}
\end{center}

\caption{\small The scalar (dashed line) and tensor (continuous line)
potentials in the IR large-$z$ TC region
for $e=0.1$, $c=13/(1+e)$, $F=1000$ (note log$_{10}$ scale). The large-$z$ tail is
given by \nr{VSlargez} with $\Xi\approx0.0014\Lambda$.
Two lowest scalar and
tensor excitations are plotted, the ordering is
$E_0^{(S)}<E_0^{(T)}<E_1^{(S)}<E_1^{(T)}$. The corresponding mass values,
$m^2\equiv E$, are in \nr{masses}.
}
\la{largz}
\end{figure}

\begin{figure}[!t]
\begin{center}

\includegraphics[width=0.49\textwidth]{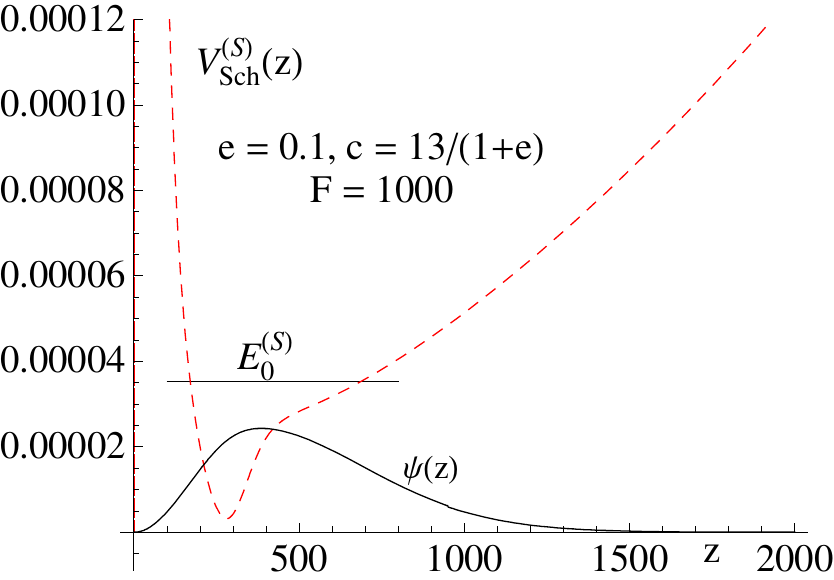}
\end{center}

\caption{\small The scalar potential
and the energy
($E_0=0.355\cdot10^{-4}$) and wave function
of the lowest scalar state 
(note linear scale). 
}
\la{wf}
\end{figure}

To have an overall picture of the situation, we first plot in Figs.~\ref{largz}
(on logarithmic scales) and \ref{smallz} (linear scales, large $V_\rmi{Sch}^{(S)}$)
the $f=1$
Schr\"odinger potentials in \nr{tensscalpot} choosing
the same reference values \nr{refval} as in the discussion of thermodynamics:
$e=0.1$, $c=13/(1+e)$, $F=1000$. Later we shall study the approach to an
IR fixed point by taking $e\to0$. Note that formally the potentials are
plotted in units of $\Lambda$,
$V_\rmi{Sch}^{(S)}/\Lambda^2$ vs $\Lambda z$, but we have chosen $\Lambda=1$.

Overall, the potentials follow very well the pattern in \nr{VSlargez} with
$\Xi\approx1/730$ as in Fig.~\ref{fields}, with some important modifications:
there appear local minima at the ETC scale and the scalar potential
has more structure than the tensor one and also binds more strongly.

\subsection{Stable states and quasinormal modes at the TC level}
The lowest energy stable states of the system are those near the minimum of
the potential $V_\rmi{Sch}$.
Computing the four lowest scalar and tensor stable states one finds that
their mass values are (as always, in units of $\Lambda=1$)
\ba
m^{(S)} &=& (0.005948, 0.0077235, 0.0090700, 0.010182)\nn
m^{(T)} &=& (0.0069018, 0.0083675, 0.0095586, 0.010581)
\la{masses}
\ea
In spite of the prominent dip in the scalar potential
the lowest scalar mass, the analogue of standard model Higgs, is thus not
particularly well separated from the higher excitations. Its wave function
is plotted in Fig.~\ref{wf} and one sees
that the narrowness of the potential well pushes the state higher. Masses
should be compared with $\Tlow=1.0\times10^{-3}\Lambda$, see Fig.~\ref{thermotc},
and one observes that the lowest mass $\approx2\pi \Tlow$.

The tensor potential is well approximated by the form
\nr{VSlargez}. In fact, one could use this to define the IR scale $\Xi$
via the equation
\be
\Xi(c,e)={(5/12)^{1/4}\over z_\rmi{min}},\quad V'(z_\rmi{min})=0.
\la{defXi}
\ee
Numerically, this is close to the alternative definition \nr{bIR},
$b(z)\to\exp(-\Xi^2z^2)$;
for $e=0.1,\,c(1+e)=13$ the former gives $\Xi=1/680$, the latter $\Xi=1/730$.

Since the approximation \nr{VSlargez} works so well,
the approximate quasinormal tensor modes can be read from
the results of subsection \ref{qnmodel} by noting that there $3\Xi^2=1$,
i.e., in Fig.~\ref{Vffig} one should replace $\omega\to\sqrt{3}\,\Xi$. For the
scalar case the quasinormal modes have not been computed.

\subsection{Stable states and quasinormal modes at the ETC level}
The $f=1$ potential over the entire range was plotted in Fig.~\ref{largz}.
A close-up of the
ETC region, $z\sim1,\,\lambda\sim1, \,V_S\sim 1$, is shown in Fig.~\ref{smallz},
left panel. Overall, the potential is confining but extremely wide, one has to
go to $z\sim10^6$ before the large $z$ part reaches the magnitude 100 of the peak
in the ETC region. Due to the confining
nature of the potential, there are stable states of large excitation number.
In fact for the potential \nr{VSlargez}, which approximates very well the potential
in Fig.~\ref{largz}, one has
\be
m^2_n=12\Xi^2 (n+2).
\la{highmass}
\ee
Thus, for $m=1$ the excitation number would be $n=44410$ and the mass
splitting $\Delta m=\sqrt{3/n}\,\Xi$. However, there also are local
minima in the potential. These arise due to the
quasiconformal behavior of the beta function
near $\lambda=1$ and the related rapid variation of the bulk fields in
Fig.~\ref{fields}.

Potentially, the most interesting feature is
that the scalar potential even dips clearly below zero, so that
even stable states are thinkable. However, for all the parameters
the dip seems to be so
narrow that the energy levels are lifted to the metastability region.

One may ask what the basic reason for the appearance of this structure in
the ETC region is. It really comes from the transition of $b(z)$ from
the UV conformal region with $b(z)=\CL_U/z$ to a quasi-conformal
region with $b(z)\approx\CL_I/z$. This is described by Eq.~\nr{zbfit}
and inserting this to $V_\rmi{Sch}$ one sees that the peak structure
is formed.

To see the effect of minima at $V>0$, forget first the increase of the potential
at very large $z$. The minima then would correspond to metastable states.
To estimate their energies,
one may demand that the wave function of the potential in Fig.~\ref{smallz}
vanish at $z=0.01$ and $z=1.2$. One finds that this potential can bind two states
with masses
\be
m_0=4.42,\quad m_1=8.92,
\la{massestimates}
\ee
shown as horizontal lines in Fig.~\ref{smallz}.
Of course, these states are only metastable. The tunneling actions for these
states are of the order of 1 so that estimates of widths are are very unreliable.
Restoring the increase of the potential at large $z$ one sees that these states
simply disappear in the dense spectrum of large $m$ states \nr{highmass},
effectively doubling two of the states \nr{highmass}. However, the levels
\nr{massestimates} will reenter when considering quasinormal modes.

\begin{figure}[!t]
\begin{center}

\includegraphics[width=0.49\textwidth]{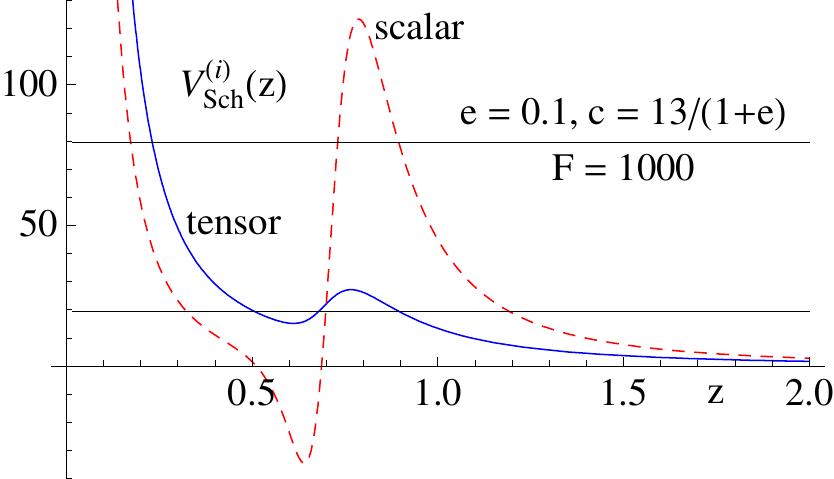}\hfill
\includegraphics[width=0.49\textwidth]{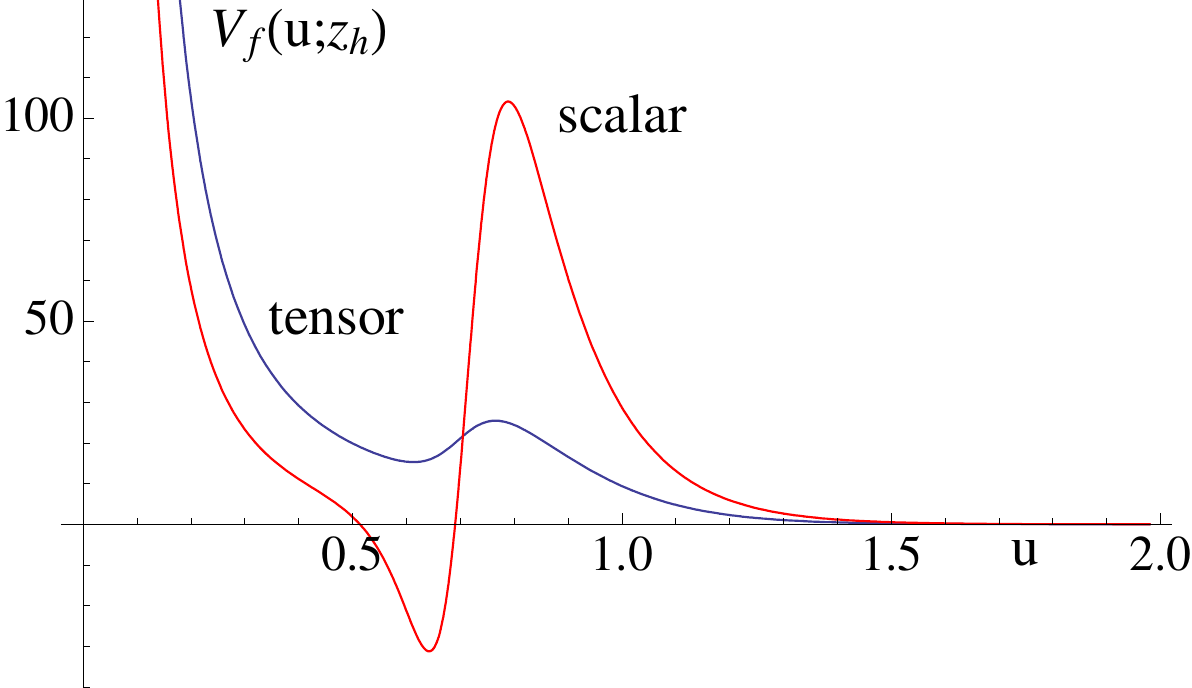}
\end{center}

\caption{\small Left panel: The scalar and tensor potentials in the UV small-$z$ ETC region
for $e=0.1$, $c=13/(1+e)$, $F=1000$
(note linear scale). For $z\to0$ both potentials approach $\sim 15/(4z^2)$. For $z\to\infty$
one has to go to $z\approx10^6$ before $V_\rmi{Sch}^{(i})$
grows back to the value $100$ (Fig.~\ref{largz}).
Right panel: The $f\not=1$ scalar and tensor potentials
$V_f^{(i)}(u;z_h)$ in \nr{Vf} for the same values of $e,c,F$ and
for $\lambda_h=\lambda_c=0.71,\,z_h=0.71, T= \Tlow=0.6\Lambda$.
}
\la{smallz}
\end{figure}

\begin{figure}[!tb]
\begin{center}

\includegraphics[width=0.49\textwidth]{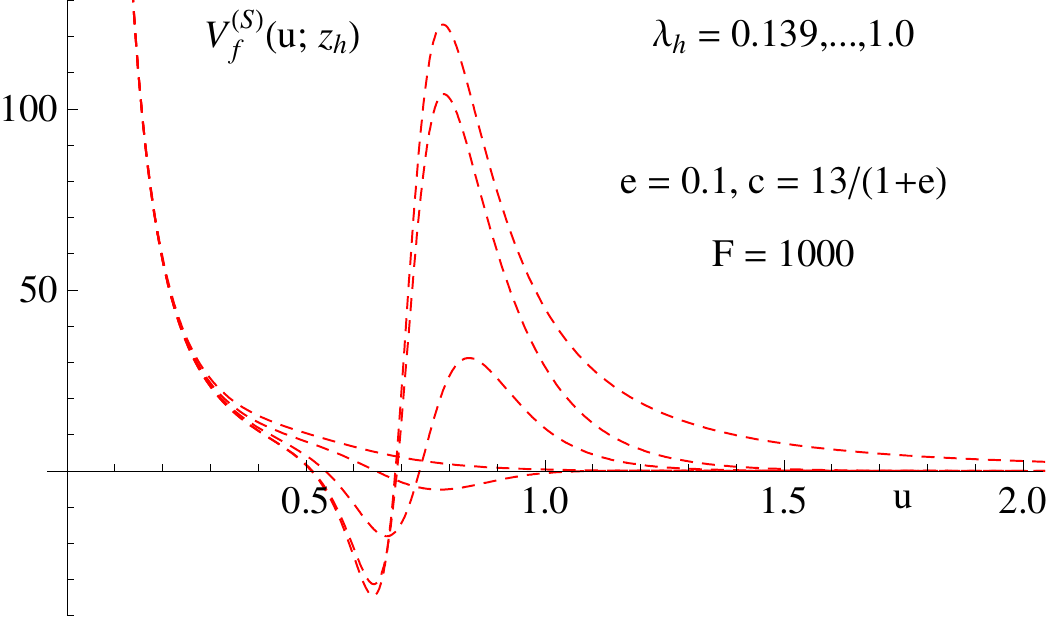}\hfill
\includegraphics[width=0.49\textwidth]{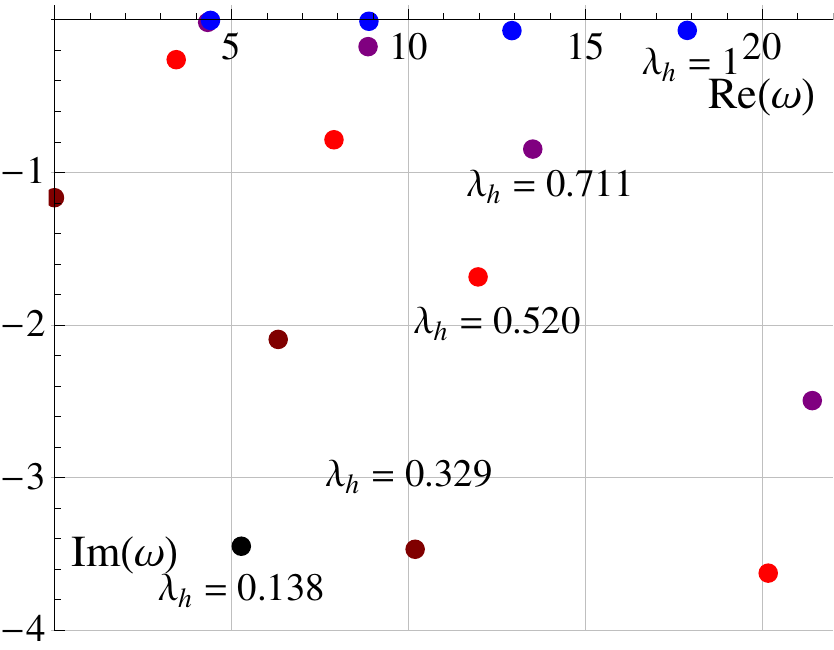}
\end{center}

\caption{\small Left panel: The dependence of the scalar potential $V_f^{(S)}(u,z_h)$ for
$\lambda_h=0.138,\,0.329,\,0.520,\,0.711,\,1.0$,
$z_h=0.57,0.719,0.724,0.92,2.96$, $\pi T\approx 1/z_h$. The peak disappears when one is
approaching the UV, when $\lambda_h$ decreases.
Right: Corresponding quasinormal modes.
}
\la{vfzh}
\end{figure}

The appropriate $f\not=1$ potentials are plotted in Figs.~\ref{smallz} and \ref{vfzh}.
Fig.~\ref{smallz} compares the $f=1$ potentials $V_\rmi{Sch}^{(i)}(z)$ and $f\not=1$
potentials $V_f^{(i)}(u;z_h)$ at $T=\Tlow$. Fig.~\ref{vfzh}, left panel, shows the $f$
dependent potential $V_f^{(i)}(u;z_h)$ for a wider range of $z_h$.
To appreciate the relation $\lambda_h=\lambda(z_h)$, see figure \ref{fields}. One sees that,
as suggested by Fig.~\ref{Vffig}, when $z_h$ approaches the UV, the
metastability peak is washed away.

The quasinormal spectrum computed for various values of $z_h$ is
also shown in Fig.~\ref{vfzh}. For large values of $z_h$
($\lambda_h\gsim1$ the metastability peak is efficient and there are states with
negligible imaginary part. These are the metastable states suggested physically by
the existence of the peak in the potential. Concretely, the lowest states
for four of the $\lambda_h$ values are
\ba
\lambda_h=1,\,z_h=2.96:&&4.41-0.0036i,\quad 8.89-0.010i,\quad 12.9-0.071i,\quad 17.9-0.069i\nn
\lambda_h=0.711,\,z_h=0.922:&&4.33-0.018i,\quad 8.86-0.18i,\quad 13.5-0.847i,\quad 21.4-2.50i\nn
\lambda_h=0.520,\,z_h=0.724:&&3.44-0.26i,\quad 7.90-0.79i,\quad 12.0-1.68i,\quad 20.2-3.62i,\nn
\lambda_h=0.329,\,z_h=0.719:&&0.00-1.17i,\quad 6.32-2.09i,\quad 10.2-3.47i,\quad 29.5-4.80i.
\la{ETCstates}
\ea
Comparing with \nr{massestimates} one sees that for $\lambda_h\gsim1$ the temperature is already
so low that the real parts are very close to the estimated energies of metastable
states. However, now one also has a controlled determination of the imaginary
parts. For increasing $\pi T\sim 1/z_h$ the
imaginary part rapidly grows, the states are melted away. The real part of the lowest state
starts decreasing and soon is even very close to zero. Note that this is quite different from what
was observed with the model computation plotted in Fig.~\ref{Vffig}. The reason is obvious:
the pattern in Fig.~\ref{vfzh} is determined by the effective vanishing of the ETC metastability
peak when $T$ grows, while there is no such effect in the TC region.

\begin{figure}[!b]
\begin{center}

\includegraphics[width=0.49\textwidth]{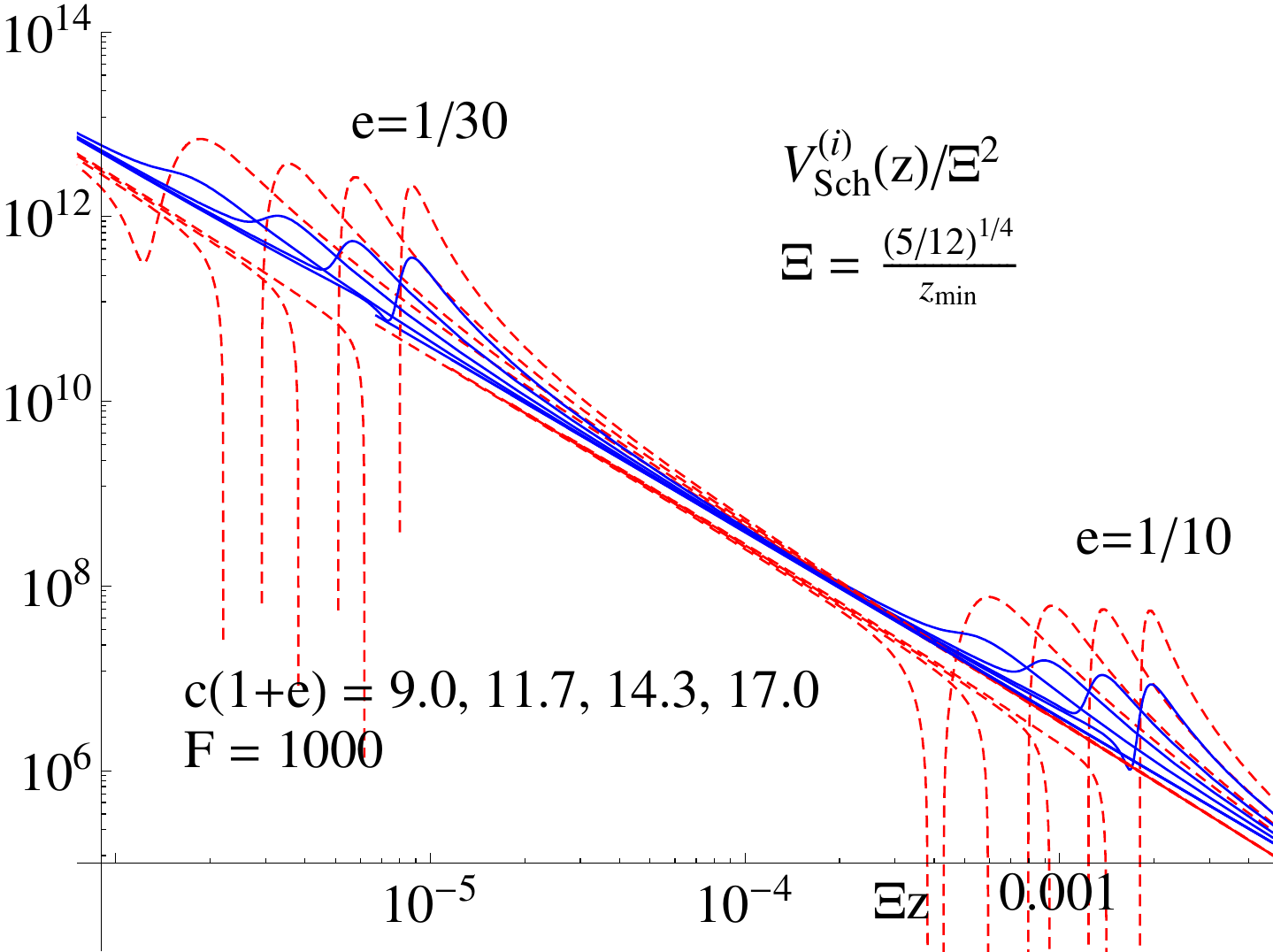}\hfill
\includegraphics[width=0.49\textwidth]{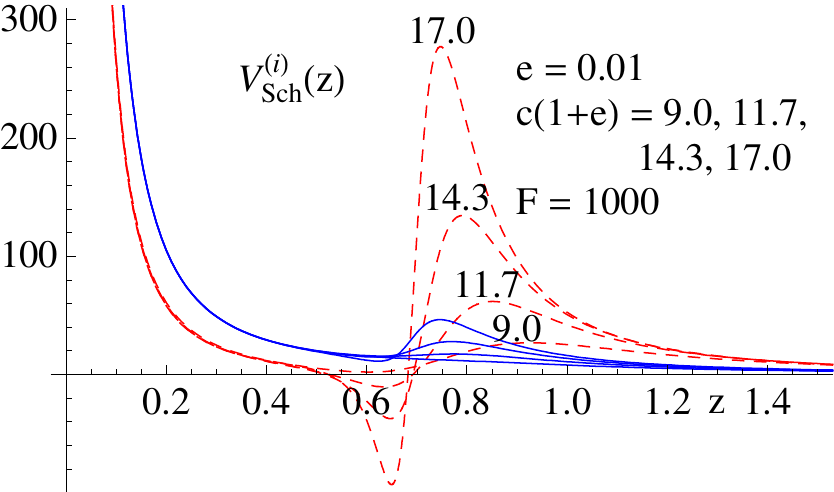}
\end{center}

\caption{\small Left panel: The Schr\"odinger potentials scaled with the
IR scale \nr{defXi} for two values
of $e$, plotted vs $\Xi z$.
Right panel: Dependence of the Schr\"odinger potentials on $c$ at fixed $e=0.01$. If one
increases $e$ to $0.1$, the height of the peak and the depth of the dip
at $c(1+e)=14.3$ increase by $\sim30\%$.
}
\la{pardep}
\end{figure}

Finally, we consider how our results are affected by the choice of the parameters
in the beta function, i.e. in the dilaton potential. First, since for $e\neq 0$ the
theory always confines at large $\lambda$, the large $z$ behavior and
the spectra at the TC level are practically independent on the choice of
parameters as long as $e\neq 0$.
Second, at small $z$, i.e. at the ETC level, there arise
structures more sensitively dependent on the values of the parameters. 
This is illustrated in
Fig.~\ref{pardep} which gives examples of how the Schr\"odinger potentials
vary when parameters are changed. In the plots, we show the potentials scaled
with the IR scale $\Xi$ defined in \nr{defXi} vs. $\Xi z$.
As we have already noted, the shape of the potential at small $z$
consists of a minimum followed by a peak as $z$ increases.
There are two main effects which arise as the parameters
in the potential are changed: The first effect is that the location of the
high temperature minimum shifts towards larger values of $z$ as
the parameter $e$ is increased. The left panel of the figure shows how
this happens for two values $e=0.1$ and 0.3. The parameter $F$ has similar
effect, i.e. larger value of $F$ implies larger separation between the high
and low temperature minima.
The second quantitative feature is the depth of the high temperature minimum
and the height of the peak. This is mainly controlled by the parameter $c$ as
can be seen from the left panel of Fig. \ref{pardep}, and in more detail for
$e=0.01$ from the right panel.
In conclusion, the qualitative shape of the Schr\"odinger potentials and
our corresponding results for the spectra appear
robust as the parameters are varied so that the walking nature of the theory
is preserved. Of course the quantitative results like the number of metastable
states at the ETC level, and how rapidly they melt away at increasing
temperature will depend on
the exact numerical values chosen for the parameters.

\subsection{Approaching the IR fixed point, $e\to0$}

When $e\to0$ in \nr{ansatz}, the $\beta$ function approaches one with an IR
fixed point at $\lambda=1$ (we remind that, in practice, always $a=\fra23 c$):
\be
\beta(\lambda)\approx -{c\over1+a}[(1-\lambda)^2+e].
\la{nearirfp}
\ee
In the limit the theory is conformal both in the UV and in the
IR. Thus one expects the mass spectrum to approach zero,
in the potential language the confining
part $+9\Xi^4z^2$ of the potential disappears. How this happens can be derived
as follows.

Integrating $d\lambda/dt=\beta(\lambda)$, $t=\log(\mu/\mu_0)$ in the approximation
\nr{nearirfp} one has
\be
\lambda(t)=1-\sqrt{e}\tan({c\over 1+a}\sqrt{e}\,t),
\ee
normalised so that $\lambda(0)=1$. This decreases monotonically from $+\infty$,
reached at the value $t<0$ defining an IR scale:
\be
t_\rmi{IR}=\log{\mu_\rmi{IR}\over \mu_0}=-{(1+a)\pi\over 2c\sqrt{e}}.
\ee
The UV scale can be defined as the value of $t>0$ for which $\lambda(t)$ has decreased to
zero:
\be
t_\rmi{UV}={(1+a)\pi\over 2c\sqrt{e}}-{1+a\over c^2}\approx -t_\rmi{IR}.
\ee
Thus
\be
{\mu_\rmi{IR}\over\mu_\rmi{UV}}=e^{t_\rmi{IR}-t_\rmi{UV}}=
\exp\left[-\left(\fr23+{1\over c}\right){\pi\over\sqrt{e}}\right],
\ee
where $a=\fra23 c$ was inserted. We thus expect that, at small $e$, the
IR scales and masses behave as
\be
\Xi\sim m=A\exp\left[-{D\over\sqrt{e}}\right],\quad
D=\left(\fr23+{1\over c}\right)\pi,
\la{qcmasses}
\ee
$A$ = constant.

This prediction works surprisingly well. All the masses are proportional to the IR scale
$\Xi=(5/12)^{1/4}/z_\rmi{min}$, defined in \nr{defXi}.
Here $z_\rmi{min}$ is the value at which the
tensor potential has its IR minimum. Its dependence on $e$ at
fixed $c(1+e)$ (keeping this fixed keeps the leading UV behavior unmodified by
the variation of $e$) is shown in Fig.~\ref{echange}, left. For masses the
behavior is very similar:
\ba
m_0^{(S)}(c(1+e)=9,e)&=&5.9874\exp[-2.4348/\sqrt{e}],\nn
m_0^{(T)}(c(1+e)=9,e)&=&6.6283\exp[-2.4337/\sqrt{e}],\nn
m_1^{(S)}(c(1+e)=9,e)&=&7.7630\exp[-2.4342/\sqrt{e}],\nn
m_0^{(S)}(c(1+e)=17,e)&=&14.671\exp[-2.2764/\sqrt{e}].
\la{massvalues}
\ea
According to \nr{qcmasses} the calculated coefficient in the exponent is
$D=7\pi/9=2.4435$ for $c=9$ and $D=37\pi/51=2.2792$ for $c=17$.

The pre-exponential factor in \nr{mKT} is non-universal. One finds that
the $c$ dependence of $\Xi$ or masses at fixed $e$ is powerlike,
$\sim c^{p(e)}$, with $p(e)=1.287/e^{0.228}$
over the interval $0.01<e<0.1$. An example at fixed $e$ is shown in Fig.~\ref{echange}, right.

Comparing \nr{massvalues} with \nr{masses} one sees that the relative hierarchy
of the states remains unchanged when $e\to0$. One could have hoped the lowest
scalar state to separate from the others in this limit \cite{dst}, but the model clearly needs
new dynamical input to accomplish this.

Since "Miransky scaling" \nr{qcmasses} generally arises as one approaches a transition into
a conformal theory, the prediction \nr{qcmasses} and its verification by the full
numerical analysis, provides another successful test for the
consistency of the gauge/gravity framework considered in this paper.

\begin{figure}[!t]
\begin{center}

\includegraphics[width=0.49\textwidth]{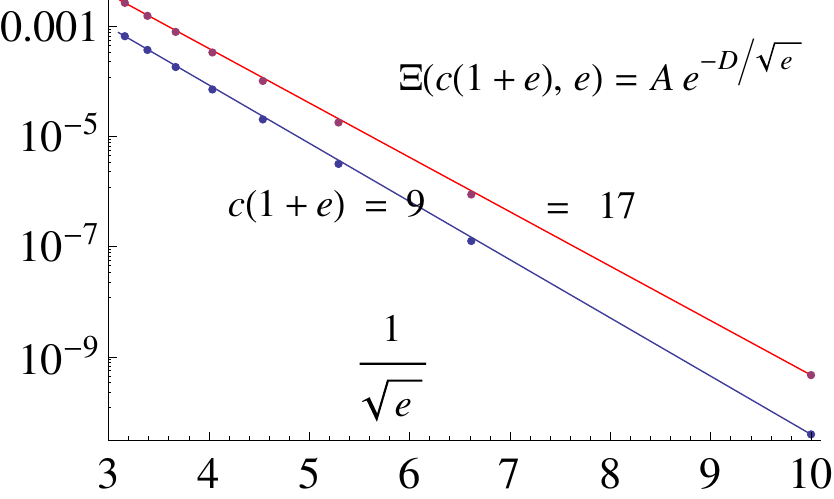}\hfill
\includegraphics[width=0.49\textwidth]{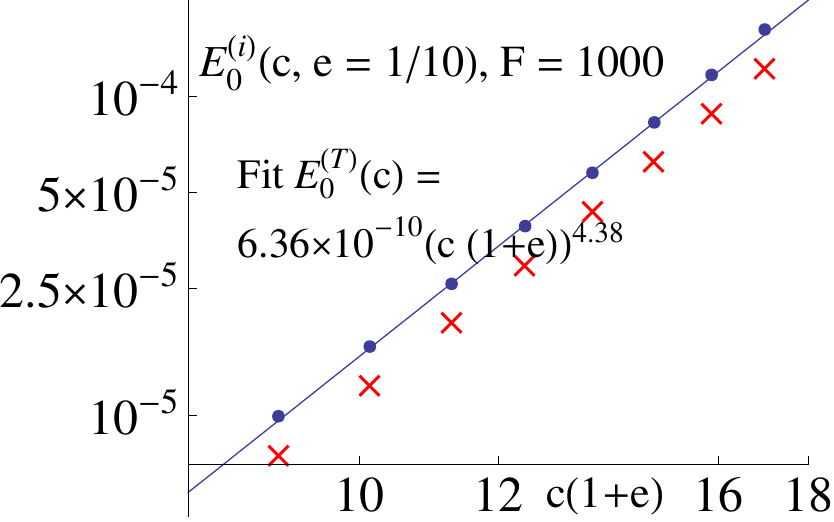}
\end{center}

\caption{\small Left panel: The dependence of the IR scale $\Xi$ defined in
Eq.~\ref{defXi}
on $e$ at fixed $c(1+e)=9\,\,(A=1.4769,\,\,D=2.4358)$
and $c(1+e)=17\,\,(A=3.60638,\,\,D=2.277)$. All masses
behave very similarly. Right panel: The dependence of the tensor and scalar (crosses)
masses squared on $c$ at fixed $e=0.1$.
}
\la{echange}
\end{figure}
\section{Conclusions}
In this paper, we have considered five dimensional dilaton gravity solutions which, in
the framework of holography, are dual to a quasi-conformal
gauge theory. In these theories the beta function almost vanishes, and the
coupling is almost constant between two widely separated energy scales.

The dilaton potential $V(\phi)$ was determined so that using it as input
to Einstein's equations, the required beta function was obtained
as output. Computing numerically solutions with and without black hole, 
we determined the thermodynamics of these theories as
well as their vacuum mass and finite temperature quasinormal spectrum.
The gravity solutions
imply that the corresponding field theory undergoes two phase transitions.
Starting from low temperatures, there
first is a transition from confining to quasi-conformal phase at $\Tlow$
and then, at significantly higher temperature $\Thigh\sim 10^3 \Tlow$
a further transition to a strongly interacting 
plasma phase.

We determined the spectra of scalar and tensor states and studied how
they are affected by quasiconformality between widely separated energy scales.
The vacuum (no black hole in dual) mass spectrum is little affected, there
are the usual vacuum states with masses $m^2\sim n\Tlow^2$ up to very large
values of $n$. However, when approaching conformality at $e=0$, the overall scale of
the mass spectrum was shown to exhibit Miransky type scaling,
$m\sim\exp\left[-D/\sqrt{e}\right]$, with known $D$. This shows the robustness
of the holographic scheme: the approach to conformality is built in the ansatz
\nr{betafn}, but the masses are computed within a rather complicated scheme of
choosing an ansatz for the bulk metric, solving the bulk fields from Einstein's
equations and solving field equations in this background. Still the Miransky
scaling comes out numerically with great accuracy. Another sign of the robustness
is that the Schr\"odinger potentials computed for scalar states contain negative
parts which, in principle, could predict stable states at both $\Thigh$ and
$\Tlow$. However, it seems that these dips never can bind any states.

In contrast to the vacuum spectrum, the quasinormal spectrum (black hole in the dual)
in strongly affected. In fact, there are new states with small imaginary part at
the $\Thigh$ scale, see \nr{ETCstates}. The existence of these states is already suggested
by dips in the vacuum potential. Dynamically the origin of these dips is in the transition
of the system from one conformal phase in the UV ($b(z)=\CL_U/z$) to a quasi-conformal
phase ($b(z)\approx\CL_I/z$).

The emphasis of this article was on quasi-conformality between two widely separated
energy scales. This is what walking technicolor is built on, but it is obvious that
for realistic applications to technicolor models, one would need
to take better into account the dynamics of flavor degrees of freedom.

\vspace{1cm}
{\it Acknowledgements}.
JA thanks the Magnus Ehrnrooth foundation and TA the Vaisala foundation for financial support.

\appendix

\section{Calculation of the Schr\"odinger potentials and their eigenvalues}
In this Appendix, we shall discuss the integration of the Einstein's equations
\nr{eka}-\nr{system} for the case $f=1$. As discussed in \cite{kiri3}, there are
different types of solutions (generic, bouncing), but only one special type
corresponds to confinement. The trick of obtaining this solution is to start numerical
integration at some very large value of $z$ or $\lambda(z)$
using as the initial condition \nr{bIR}.
This automatically leads to the correct special solution. Most concretely this is
seen from the large $\lambda$ limit \nr{betaIR} of the beta function; the unwanted
solutions have $\beta\to-3\lambda$.

Following Eqs.~\nr{bIR} - \nr{largez} we start the integration of \nr{eka} - \nr{system}
with $f = 1$ using the initial conditions
\ba
b(z_i) &=& b(\lambda_i) =
b_0\,\lambda_i^{-2/3}(\fra23\log\lambda_i)^{1/2},\\
W(z_i) &=& W(\lambda_i) = \fra{1}{3}\sqrt{V(\lambda_i)}\\
\lambda(z_i) &=& \lambda_i,
\ea
where $b_0$ is an arbitrary constant. Now choose some numerical
value for $b_0$, $z_i$ and $\lambda_i$ (say, $b_0=1$, $z_i=40$, $\lambda_i=10^{200}$),
and integrate the system of differential equations
towards smaller $z$ until the $b \sim 1/(z-z_{UV})$ divergence is encountered
at the AdS boundary. Then the energy scale may be readjusted to a chosen value
of $\Lambda$ (as in \nr{la2loop}) and the coordinate origin moved to $z_{UV} = 0$ by the scalings
described in appendix A of \cite{kiri4} (since $f = 1$, only scalings 2. and 3.
are needed). This eliminates the effect of the arbitrarily chosen constants $b_0$ and $z_i$.

Note that since the approximation error is $\CO(1/\log\lambda)$, $\lambda_i$ has
to be very large. It also has to be very large in order to reach the large-$z$
minimum $z_\rmi{min}$ of the potential, due to the extremely slow increase of $z(\lambda)$.

After this procedure we have a set of functions $b(z),\, \lambda(z)$ for each set of
parameters $c,\,e$, $z$ ranging over the interval $(0, z(\lambda=10^{200}))$.
Given this set, the potential $V_\rmi{Sch}(z)$ can be computed from \nr{tensscalpot}.

Once the potential has been computed, a number of lowest states of the eigenvalue
spectrum can be solved using the standard shooting method. Let us choose as
a reference case $c = 9/(1+e), \,e = 1/100$, for which
$z_\rmi{min,\,ref}= 2.06089 \cdot 10^{10}$.
We set in the reference case the boundary conditions
$\psi(z_{l,r}) = 0$ at $z_l = 10^9$ and at
$z_r = 5\cdot 10^{11}$, far above and below the minimum. For other values
of $c,\,e$, we simply scale
$z_{l,r}\rightarrow z_{l,r} z_\rmi{min}/z_\rmi{min,ref}$.
We then solve by numerical rootfinding for values
of $m^2$ where the numerically computed solutions with the two boundary
conditions can be matched such that they are continuously differentiable.


\section{Numerical evaluation of the quasinormal modes}

The quasinormal modes can be numerically evaluated from the Schr\"odinger
type equation \eqref{seq}; for a review, see \cite{starinets_qn}. 
Since the equation is linear, given a value
of $\omega$ a solution is completely determined by two complex constants. One of them is
simply a normalization, and the other can be
determined from the in-going wave condition at the horizon, which in terms of
$u$ can be expressed as
\begin{equation}
  \psi(u) \rightarrow e^{+i\omega u}\quad \text{for large $u$}.
\end{equation}
This reduces the problem to finding those values of $\omega$ where the solution satisfying
this condition is normalizable, i.e. $\phi(z) \rightarrow z^4$ when $z\approx u\rightarrow 0$.

The standard method, used in Section \ref{qnmodel}, requires expanding the solutions
around $z=0$. However, in the present gravity dual these expansions are not power series
in $z$ but contain logarithmic terms, see for example \nr{la2loop}. To avoid this
we apply a more direct numerical method of removing the unnormalisable component from
the solution.

In practice, since $\omega$ is never numerically exactly at a quasinormal mode, we expect
that the solution will contain some part of the non-normalizable $\phi \rightarrow 1$
solution at the boundary, and $\psi(u) \rightarrow u^{-3/2}$ will therefore diverge.
If we are close to a quasinormal mode, it seems intuitive that this divergence will
happen only when $u$ is very small (this is the heuristic of the algorithm, since we
do not have a proof of this). This suggests a way to look for the quasinormal modes:
given a trial $\omega$, compute numerically the solution from large $u$ towards the
boundary and find the smallest-$u$ local minimum of $\lvert(\psi(u)\rvert$. Denoting
its position by $u_{min}$, the zeros of $u_{min}(\omega)$ are at the quasinormal modes.
With finite precision numerics we can then simply look for the minima of $u_{min}(\omega)$.
See Fig.~\ref{umins} for a visualization of the process when applied to the model of
Section~\ref{qnmodel}, for which the usual power expansion method could be applied.

\begin{figure}[!b]
\begin{center}
\label{umins}
\includegraphics[width=0.49\textwidth]{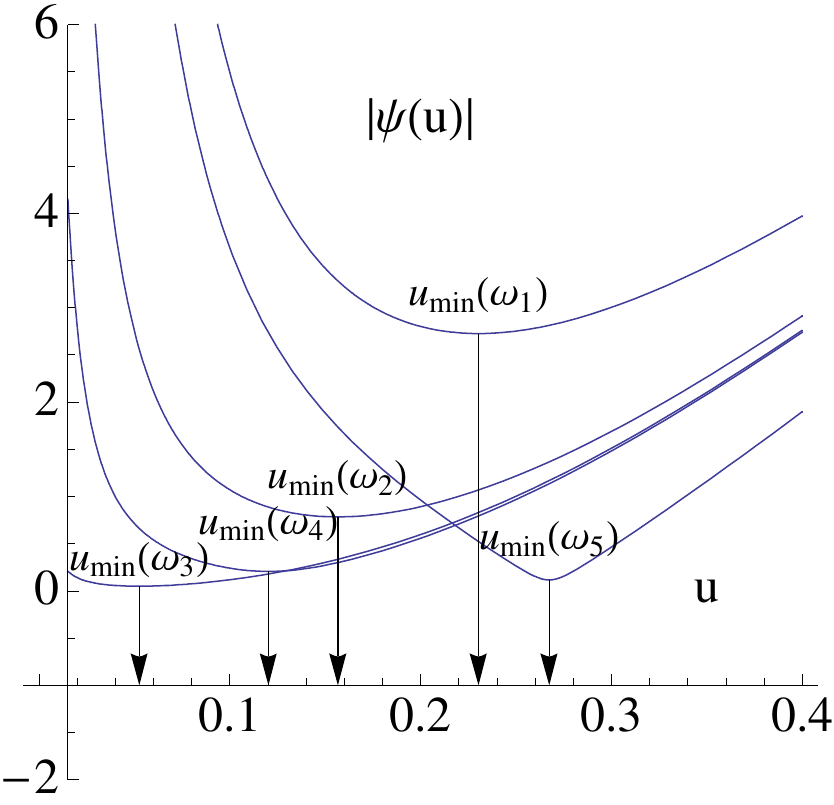}\hfill
\includegraphics[width=0.49\textwidth]{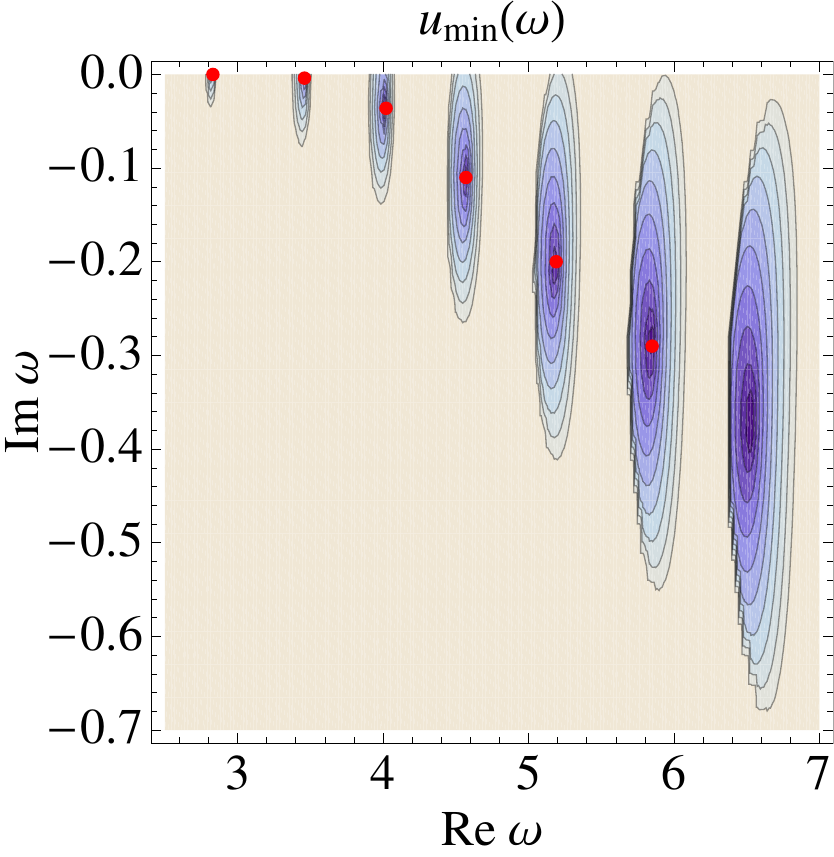}
\caption{\small The numerical method applied to the analytic model \eqref{bz} with
$z_h = 4$. Left panel: An example of looking for the fourth minimum. The plot shows
$|\psi(u;\omega)|$ for $\omega_n \in \{4.550- 0.1000i, 4.573- 0.1083i, 4.575- 0.1137i, 4.576- 0.1125i,
 4.597- 0.1167i\}$. The arrows mark the location of the minima $u_{min}(\omega)$.
 Notice how the minima first approach the origin, then after the closest point
 $\omega_3 = 4.575- 0.1137i$ they again start to move away. This is the
 computed value of the quasinormal mode. Right panel:
 The minima $u_{min}(\omega)$ on the complex plane. On the gray area
 $u_{min} > 0.5$, and $u_{min}$ decreases for shades towards dark blue.
 The red dots show the positions calculated in fig.~\ref{Vffig} for comparison.}
\end{center}
\end{figure}


To start the integration of \nr{seq} one may note Eq.~\nr{Vflimit}, at
large $u$
\begin{equation}
V_f^{(i)}(u;z_h)\to A^{(i)}(z_h) e^{\dot f_hu}\to 0,
\end{equation}
where $A^{(i)}(z_h)$ is as defined in Eq.~\nr{VflimitAfactor}.
With this potential \nr{seq} has the analytic solution
\be
\psi(u)=
C_1 I_\nu\left({2\over-\dot f_h}\sqrt{A^{(i)}(z_h)}\,e^{\fra12\dot f_hu}\right)+
C_2I_{-\nu}\left({2\over-\dot f_h}\sqrt{A^{(i)}(z_h)}\,e^{\fra12\dot f_hu}\right),
\quad \nu={2i\omega\over\dot f_h}.
\ee
where the first term has the correct large $u$ limit $e^{+i\omega u}$. One can then
use this evaluated at some $u$ somewhat larger than $z_h$ as the initial condition of
numerical integration.
This allows us to choose the solution with the correct asymptotic at infinity,
 which is the first term, and use the solution for the exponential potential to
 start the numerics at a somewhat smaller u. With this improvement, the algorithm
 works in the analytic model of Section~\ref{qnmodel}
 for $z_h = 1\ldots4$, although at $z_h = 1$ there still
 seems to be a tendency for the minima to lie at a too small $\im \omega$.

\edoc